# Comparative Analysis of LSTM, GRU, and Transformer Models for Stock Price Prediction

Jue Xiao[1]*
Independent Researcher
The School of Business at University of Connecticut
Jersey City，NJ，07302， USA
* Corresponding author: jue.xiao@uconn.edu

Shuochen Bi[2]
Independent Researcher
D'Amore-McKim School of Business at Northeastern University
Boston，MA，02115，USA
bi.shu@northeastern.edu

Tingting Deng[1,2]
Independent Researcher
Simon Business School at University of Rochester
Chantilly, VA 20151, USA
dengtin1@gmail.com

***Abstract*—In recent fast-paced financial markets, investors constantly seek ways to gain an edge and make informed decisions. Although achieving perfect accuracy in stock price predictions remains elusive, artificial intelligence (AI) advancements have significantly enhanced our ability to analyze historical data and identify potential trends. This paper takes AI-driven stock price trend prediction as the core research, makes a model training data set of famous Tesla cars from 2015 to 2024, and compares LSTM, GRU, and Transformer Models. The analysis is more consistent with the model of stock trend prediction, and the experimental results show that the accuracy of the LSTM model is 94%. These methods ultimately allow investors to make more informed decisions and gain a clearer insight into market behaviours.**

## I. INTRODUCTION

The financial markets are integral to global economic progress, and the stock market's increasing significance in the economic domain has garnered growing public interest. The Efficient Market Hypothesis (EMH) provides a framework for understanding how financial assets are priced and why stock market fluctuations occur. This theory posits that rational investors will promptly and logically process all available market information in an ideal stock market characterised by regulation, transparency, and competition. As a result, stock prices are expected to reflect all pertinent information about a company's current and future value accurately and timely. This complexity necessitates using machine learning techniques for tasks such as forecasting stock trends, predicting stock prices, managing investment portfolios, and developing trading strategies.

Given the inherent uncertainty and variability of the stock market, accurately predicting market trends is challenging. The unpredictability stems from various factors, such as economic indicators, geopolitical events, market sentiment, and investor behaviour, which complicate modelling stock price movements. Advanced machine learning algorithms can address these challenges by processing extensive historical and real-time data, uncovering patterns, and providing more precise forecasts than traditional methods. Leveraging these sophisticated techniques can enhance predictive accuracy and make more informed decisions in the dynamic stock market environment.

Historically, forecasting models such as decision trees and support vector machines (SVM) have been employed to predict stock market trends. Decision trees offer a straightforward approach by creating a model that splits data into branches based on feature values, providing a clear path to decision-making. However, these traditional models often struggle with financial data's high dimensionality and non-linearity. As a result, machine learning techniques have evolved to incorporate more sophisticated models, such as deep learning networks and ensemble methods, better to capture the intricate relationships within stock market data and enhance predictive performance.

As deep learning models continue to evolve, stock market forecasting has transitioned from conventional approaches to cutting-edge deep learning methodologies.

Contemporary methods now encompass Recurrent Neural Networks (RNNs), Long Short-Term Memory (LSTM) networks, Gated Recurrent Units (GRUs), Graph Neural Networks (GNNs), and Convolutional Neural Networks (CNNs). Recently, there has been a growing interest in utilizing Transformer-based models and Reinforcement Learning (RL) techniques for predicting stock market trends. However, numerous surveys on stock market forecasting have limitations, such as focusing on outdated technologies and using vague classifications for models.

Previous research has tackled numerous challenges and unresolved issues in stock market prediction, leading to notable advancements. This survey addresses remaining gaps by offering a detailed review of current techniques and emerging trends. It draws from recent high-quality conference papers to comprehensively analyse advancements in LSTM, Transformer, GRU, and other relevant technologies. The goal is to provide a clear understanding of these deep learning technologies and to identify the most promising directions for future research. This approach ensures that researchers, practitioners, and educators are well-informed about the latest developments and can explore new opportunities in stock market forecasting.

## II. Related Work

### A. *Traditional stock forecasting*

Stock price forecasting seeks to predict future price movements using historical data and various indicators, aiding investors in making informed decisions about buying, selling, or trading stocks. Early investigations, such as those by Li and Ma, focused on leveraging artificial neural networks to predict stock prices and pricing options. Meanwhile, Billings et al. comprehensively evaluated different classification models, including Random Forest, AdaBoost, and Kernel Factory. Cavalcante et al. examined machine learning algorithms from 2009 to 2015 in a historical analysis. Their study discussed the application of these algorithms in several areas, including financial data processing, where they help in analysing and interpreting large datasets; trend prediction, where they are used to forecast future market movements based on historical data; and text mining, where they assist in extracting meaningful patterns and insights from textual information. This comprehensive review highlights the evolving role of machine learning in processing and understanding financial and textual data. They also addressed the challenges and unresolved issues in the field.

Recent advancements have shifted focus to deep learning techniques like CNNs, RNNs, and GNNs. This study also examined data and code availability, underscoring the progress made in deep learning methods for stock forecasting. Thakkar and Chaudhari reviewed neural network methods from 2017 to 2020, identifying needs, challenges, and future directions. Additionally, Kumbure et al. conducted a literature review of 138 articles from 2000 to 2019, categorising deep learning methods into supervised and unsupervised approaches and focusing on features and unique variables within datasets.

### B. *Stock market prediction in machine learning*

Stock market prediction involves several key tasks: forecasting stock prices, predicting stock trends, and developing trading strategies. Stock price forecasting uses time series data to predict future values and identify profitable opportunities influenced by market psychology. By analysing price changes, stock trend forecasting classifies trends into uptrends, downtrends, and sideways movements. These include the investor's approach, the size of the market capitalization, the use of technical indicators, fundamental analysis, diversification of the portfolio, risk tolerance, and the application of leverage. Each factor is crucial in shaping effective trading decisions and managing investment risk.

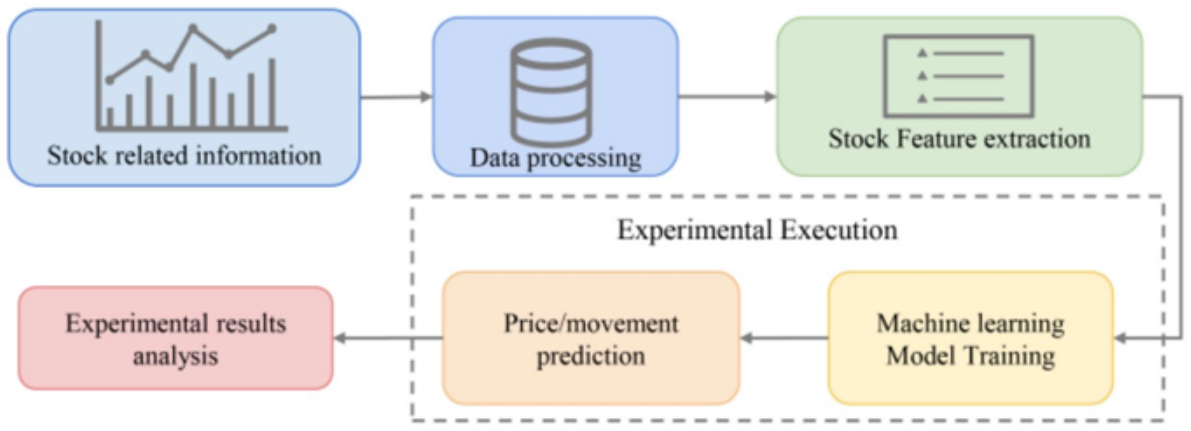

Figure 1. The Processing Framework.

Figure 1 outlines the prediction process for understanding deep learning-based stock market strategies. The process begins with data processing, incorporating stock data, charts, and text. Relevant features are selected and fed into deep-learning models for training. The final step involves analysing the model's experimental results to evaluate its performance. Common deep learning-based stock market prediction strategies include event-driven, data-driven, and strategy optimisation approaches.

### C. *A model based on recurrent neural networks*

RNNs are effective for processing sequential data, such as stock market time series, due to their ability to model historical data. Nonetheless, dealing with vanishing gradients in long-term dependencies remains a challenge. To tackle this issue, models such as LSTM, GRU, and Bi-LSTM have been introduced. LSTM incorporates gates—update, forget, and output—to regulate short-term and long-term memory. On the other hand, GRU streamlines this approach by merging the input and forget gates into a single update gate and adding a reset gate to manage information retention. These enhancements have led to significant improvements in stock market forecasting.

Time series data is frequently utilized in stock forecasting studies, yet time dependence is often neglected. To address this, Zhao et al. introduced a time-weighted LSTM model that assigns different weights to data points according to their proximity in time to the target prediction. This hybrid model, combining LSTM with other approaches, aims to enhance forecasting performance by considering temporal relevance.

### *D. Transformer model*

CNNs are highly effective for processing spatial data, such as images, by generating internal representations of two-dimensional information. In contrast, RNNs excel in handling temporal or sequential data, making them well-suited for analyzing financial news, tweets, and stock price time series.

With their ability to capture long-term dependencies, transformers are particularly useful for addressing time-dependent issues in financial data. For instance, Li et al. introduced a Transformer encoder attention (TEA) framework that leverages the attention mechanism to uncover time-related patterns in stock prices and their connections to social media texts. Additionally, models based on the traditional Transformer algorithm, enhanced with multi-scale Gaussian priors and local optimization, have been developed to manage high-frequency financial data and prevent redundant learning. These approaches include tools like the Trading Gap Splitter for Transformer, designed to learn the structural layers of financial data efficiently.

Transformer-based models are increasingly used for sentiment analysis of stock-related news. By analyzing text information, these models aim to predict market reactions based on the underlying sentiment conveyed in financial news media.

## III. METHODOLOGY AND EXPERIMENT

Time series analysis involves studying sequences of data points ordered in time, aiming to uncover patterns, trends, and make predictions. This analysis is crucial in fields like finance, where understanding historical data can inform future decisions.

This notebook delves into nine years of stock data from Tesla, one of the fastest-growing companies globally. Through this analysis, you will learn several key techniques:

1. Exploratory Data Analysis (EDA) techniques to understand the data's structure and patterns.

2. Plotting basic charts using Plotly for interactive visualization.

3. Utilizing advanced deep learning models such as GRU, LSTM, and Transformer neural networks for time series forecasting.

4. Forecasting Tesla's stock price for 30 days using these models.

This project aims to equip you with practical skills in handling and analyzing time series data, leveraging both traditional statistical methods and cutting-edge neural network architectures.

### *A. Dataset*

The data set used in this study comes from Tesla's official website, which covers product information, sales data, and user reviews published in multiple years. This data includes vehicle models, configuration details, price changes, user feedback, and other relevant market indicators. By using historical data from Tesla's website, we can deeply analyze the company's product trends and market reactions, thereby improving the accuracy and reliability of the predictive model. To ensure the integrity and representability of the data, we thoroughly cleaned and pre-processed the data, dealt with missing values and outliers, and carried out the necessary data enhancement.

**Table 1.** Check the Head and Tail of the dataset

| Unnamed: | Date | Open | High | Low | Close | Volume |
|---|---|---|---|---|---|---|
| 0 | 2015/1/2 | 14.858 | 14.883333 | 14.217333 | 14.620667 | 71466000 |
| 1 | 2015/1/5 | 14.303333 | 14.433333 | 13.810667 | 14.006 | 80527500 |
| 2 | 2015/1/6 | 14.004 | 14.28 | 13.614 | 14.085333 | 93928500 |
| 3 | 2015/1/7 | 14.223333 | 14.318667 | 13.985333 | 14.063333 | 44526000 |
| 4 | 2015/1/8 | 14.187333 | 14.253333 | 14.000667 | 14.041333 | 51637500 |

The dataset spans a total of 2274 days from January 1, 2015, to January 16, 2024, providing a comprehensive view of Tesla's stock performance over almost five years. During this period, monthly averages of opening and closing prices reveal fluctuations across different months. February consistently showed higher prices, while September and October exhibited comparatively lower averages. The bar chart visually represents this analysis, highlighting these monthly trends in Tesla's stock prices.

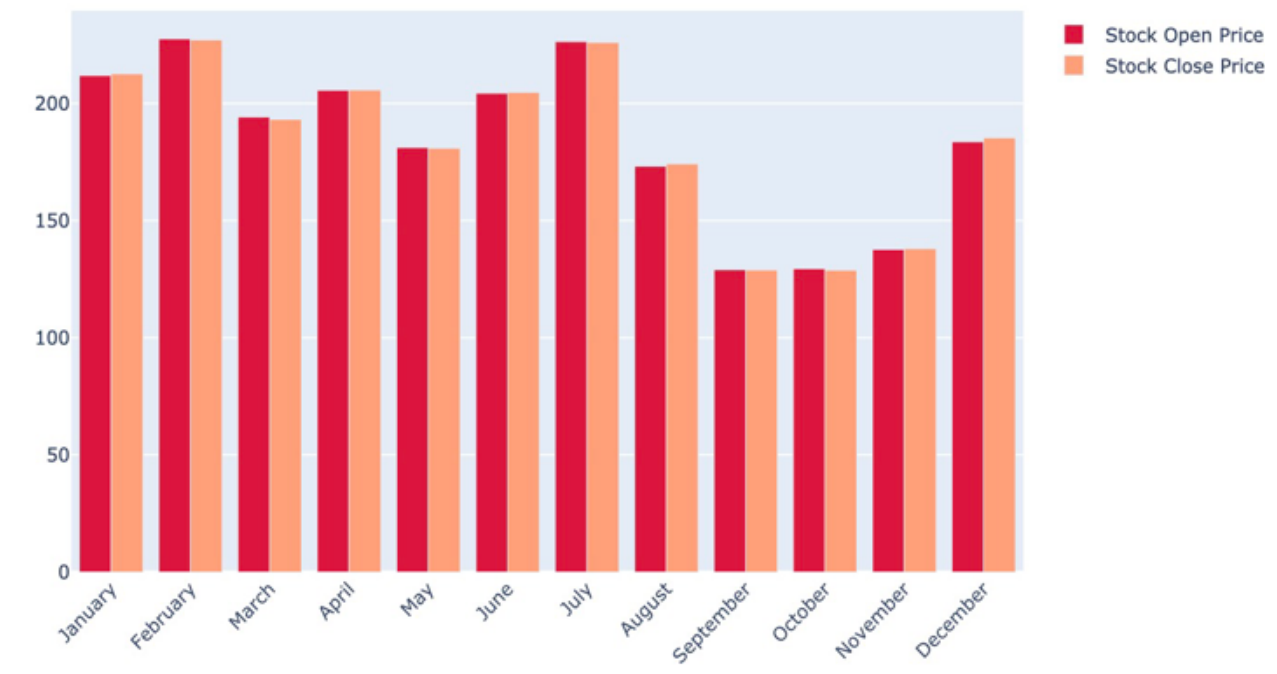

**Figure 2.** Monthwise comparison between Stock open and close price

The time series analysis of TESLA stock data spans from 2015 to 2024（Figure 2）. Initially, exploratory data analysis (EDA) revealed key insights: the dataset covers 2,274 trading days with no missing values. Visualizations depicted monthly comparisons of opening and closing prices, showcasing variations across different months. Further plots detailed the overall stock price action and volume trends over time, highlighting significant fluctuations and trends. Statistical tests like the Augmented Dickey-Fuller (ADF) were also conducted to assess stationarity. For instance, the ADF test on monthly mean high prices yielded a p-value of 0.703, indicating non-stationarity. Differencing the series improved stationarity (ADF test p-value: 0.142), enhancing its suitability for modelling.

**Table 2.** Summary of Time Series Analysis Statistics and Tests

| Statistic | Value |
|---|---|
| Total Number of Days | 2274 |
| Missing Values | None |
| Start Date | 2015/1/1 |
| End Date | 2024/1/16 |
| Monthly ADF Test (High Prices) | Test Statistic: -1.129 |
| | p-value: 0.703 |
| Monthly ADF Test (Differenced) | Test Statistic: -2.399 |
| | p-value: 0.142 |

*B. Creating Models*

1. Long Term Memory Network (LSTM)

A short-term memory network (LSTM) is a special recurrent neural network (RNN) specifically designed to process and predict time series data. Compared with standard RNNS, LSTM captures and remembers long-term dependencies effectively through a gating mechanism suitable for modeling and predicting long sequences.

LSTM model formula is as follows:

**1.Forget Gate**: This gate controls which information from the previous cell state should be discarded. It is calculated using the formula:

$$ft=\sigma(Wf\cdot[ht-1,xt]+bf) \quad (1)$$

where ft is the forget gate's output, Wf is the weight matrix, ht−1 is the previous hidden state, xt is the input at time tt, and bf is the bias term. The sigmoid function σσ determines which values to keep or discard.

2. **Input Gate**: This gate regulates the addition of new information to the cell state. It is computed as:

$$it=\sigma(Wi\cdot[ht-1, xt]+bi). \ (2)$$

where iti is the input gate's output, WiWi is the weight matrix, bi is the bias term:

3. **Cell State Update**: The cell state Ct is updated based on the previous cell state Ct−1and the new candidate values C~t:

$$C\sim t=\tanh (WC\cdot[ht-1, xt]+bC). \ (3)$$

$$Ct=ft\cdot Ct-1+it\cdot C\sim t \quad (4)$$

are the corresponding weight matrix and bias term. The tanh f() function ensures that the new cell state values are in a specific range. where C~t is the candidate cell state, and WC and bC

4. **Output Gate**: This gate determines the output of the LSTM unit based on the updated cell state. It is given by:

$$ot=\sigma(Wo\cdot[ht-1, xt]+bo) \ (5)$$

$$ht=ot\cdot \tanh (Ct) \ (6)$$

where ot is the output gate's output, WoWo is the weight matrix, bobo is the bias term, and ht is the hidden state output.

2. Gated Recurrent Unit (GRU)

Gated loop units (GRUs) are another type of recurrent neural network for processing sequence data, similar to LSTM, but with a simpler structure. It controls the flow of information by resetting and updating doors, while also having the ability to capture long-term dependencies.

Workflow for creating a model:

- Data preparation: Prepare a sequence data set suitable for the GRU and perform pre-processing such as normalization or normalization.
- Model architecture design: Design the number of layers of GRU cells and the number of cells in each layer. Select the appropriate activation function (such as tanh), loss function (such as cross-entropy), optimiser (such as Adam), and so on.

- Model training: Model training using training data sets, adjusting model parameters through backpropagation and optimisers to minimize loss functions.
- Model evaluation and tuning: Use validation sets to evaluate model performance and adjust hyperparameters (e.g., learning rate, number of hidden units) to improve model generalisation.
- Model deployment: Use test sets to evaluate final model performance and deploy models in real-world applications.

3. Transformer model

Transformer is an attention-mechanism-based neural network architecture widely used for sequence-to-sequence tasks such as machine translation, text generation, and speech recognition. It captures the global dependencies of the input sequence through a self-attention mechanism and avoids the sequential computation problem of traditional RNNS.

Workflow for creating a model:

- Data preparation: Prepare serial data sets suitable for Transformer, such as text, speech or time series data, and standardize or normalize them.
- Model architecture design: Design the encoder and decoder structure of the Transformer, including the multi-head attention layer, feedforward neural network and residual connection. Select the appropriate activation function (such as ReLU), loss function (such as cross entropy), optimizer (such as Adam), etc.
- Model training: Model training using training data sets, adjusting model parameters through backpropagation and optimizers to minimize loss functions.

# IV. Model Evaluation result

## A. *Verification and comparison of model results*

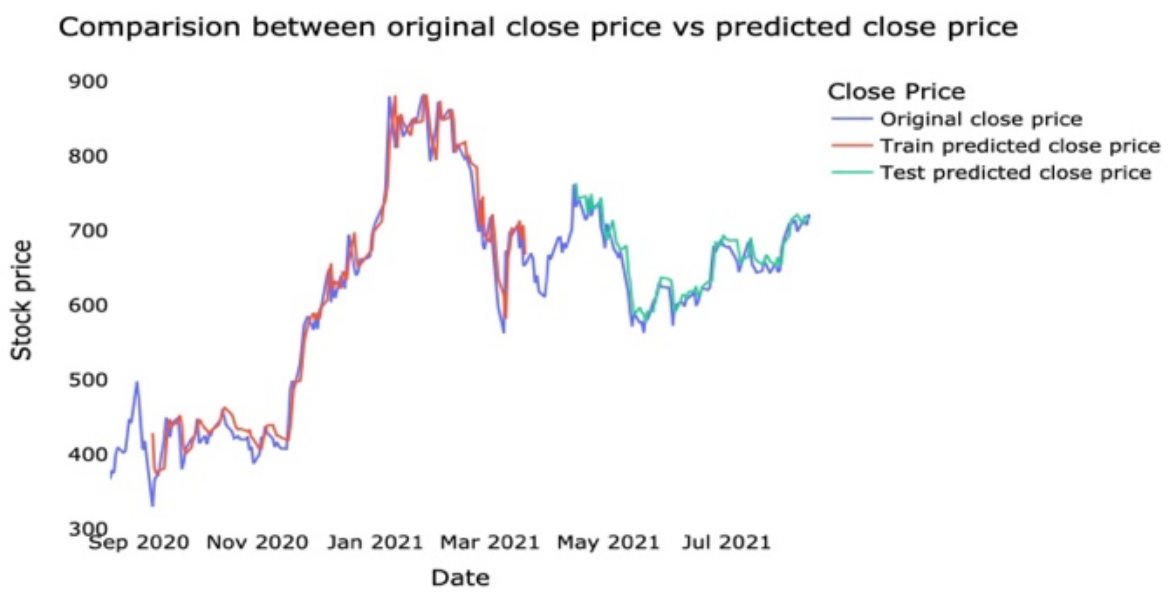

**Figure 3.** Comparison between original close price vs predicted close price

Based on the (figure 3)comparison between the original closing price and the predicted closing price of the data in this paper, Tesla's stock price forecast for the next thirty days is made by using three models, and the results are as follows：

Table 2. Comparison results of three different model predicto

| | LSTM | GRU | Transformer |
|---|---|---|---|
| R2 | 0.9802378 | 0.848738795 | 0.8027389 |
| MAE | 12.7865308 | 14.73077754 | 16.29736424 |
| MSE | 260.8710984 | 339.9002233 | 360.2837729 |
| RMSE | 26.0917386 | 18.43638314 | 16.3973472 |

In order to further verify the stock price prediction ability of the built model, this paper conducted experimental research on the stock price prediction of Tesla stock in the next 30 days. The LSTM, GRU and Transformer models are compared and analyzed. It can be seen from the evaluation index result table obtained from the comparison experiment (Table 2) that the Transformer model has the worst prediction effect, which can be intuitively seen from the table. Compared with the Transformer model, the LSTM model is closer to the changing trend of stock price and is more suitable for the stock time series. From the perspective of overall indicators, the LSTM model is more suitable for stock time series

Compared with the GRU neural network, the fitting degree of the Transformer model reaches 94.4%, which is 4.5% higher than that of the GRU neural network. Due to the addition of a bidirectional time layer on the basis of the GRU model, the fitting degree of the Transformer model reaches 90.8% in the experiment of Tesla stock price prediction for the next 30 days. Compared with the GRU model, the MAE value, MSE value and RMSE value are 14.73077754, 339.9002233 and 18.43638314, respectively, and the prediction error score of GRU model is significantly reduced.

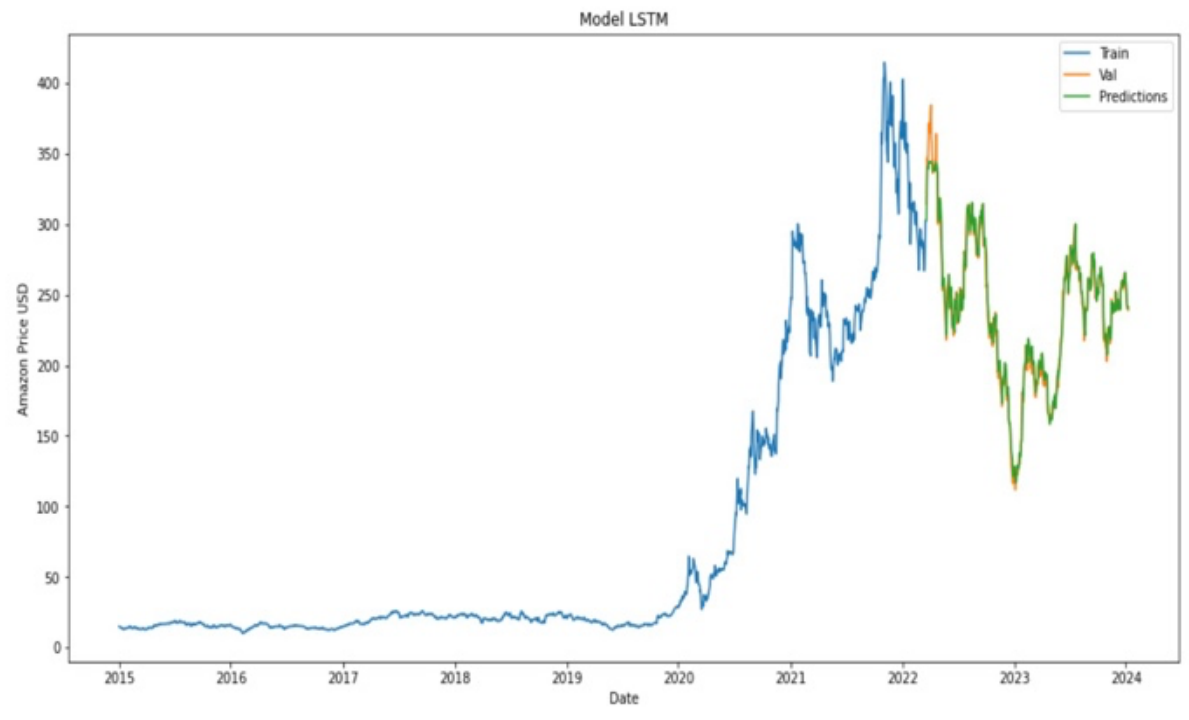

**Figure 4.** LSTM model training results

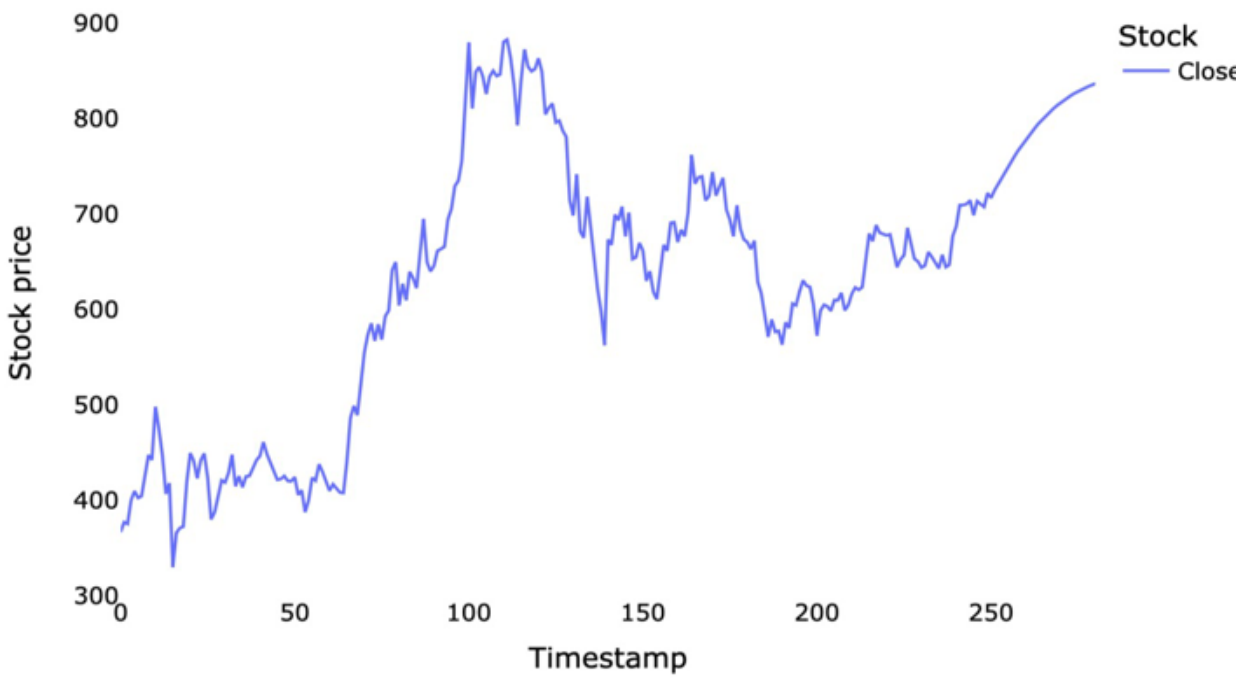

Figure 4. LSTM model forecast results for the next 30 days

Finally, it can be concluded that in the process of model prediction of stock price trend, the LSTM model with 94% accuracy was finally selected in this experiment. It can also be seen from the prediction effect in Figure 4 and Figure 5 that the LSTM model is more suitable for the prediction analysis of stock price time series than other models, which verifies the superiority of the LSTM model as a basic model.

Serial memory capability: LSTM effectively captures and remembers long-term time dependencies through its gating mechanism, which is particularly important in the non-linear and dynamic analysis of financial data.

Adaptability and generalization ability: The LSTM model shows good generalization ability in different stock price prediction tasks and can adapt to data characteristics in different time periods and market conditions.

In conclusion, the LSTM model in the Tesla stock prediction experiment has shown its strong ability in financial data analysis and prediction, especially in terms of high accuracy, sequence memory, and application flexibility. These characteristics make it an effective tool to understand and predict the stock price trend, which has important practical significance for investment decisions.

## V. Conclusion

In this paper, we explore the application of AI to stock price trend prediction, especially empirical analysis based on Long short-term memory network (LSTM), gated cycle unit (GRU) and Transformer models. Through model training and comparison of Tesla stock data, we found that the LSTM model performed well in the prediction accuracy rate, reaching 94%. The application of these technologies can not only help investors make better investment decisions but also deeply understand the dynamic changes in the market and the influencing factors behind it.

In the future, with the further development of deep learning techniques and the increase in financial market data, we expect stock market predictions to become more accurate and reliable. While there are still challenges, such as data quality, model complexity and market uncertainty, we see great potential in risk management and optimization of trading strategies through these advanced technologies. By continuously improving model design and data preprocessing methods, we can improve predictive models' performance and robustness.

In conclusion, the application of artificial intelligence technology in financial markets has shown significant advantages, especially in the field of stock price prediction. These technologies can not only provide investors with more decision support, but also facilitate broader innovation and development in the financial sector. As our understanding of data and models continues to deepen, we look forward to seeing more applications and advances in deep learning techniques in financial markets in the future.